\documentstyle[sprocl,epsf]{article}

\def\Journal#1#2#3#4{{#1} {\bf #2}, #3 (#4)}

\def\NCA{{\em Nuovo Cimento} A}

\def\NPB{{\em Nucl. Phys.} B}
\def\PLB{{\em Phys. Lett.}  B}

\def\PRD{{\em Phys. Rev.} D}
\def\ZPC{{\em Z. Phys.} C}

\def\be{\begin{equation}}
\def\ee{\end{equation}}
\def\bea{\begin{eqnarray}}
\def\eea{\end{eqnarray}}
\newcommand{\oh}{\frac{1}{2}}
\newcommand{\One}{1\kern-4.5pt1}
\newcommand{\epsfaxhax}[2]{
        \centerline{
          \hspace{-20pt}
          \epsfxsize=160pt
          {\epsfbox{#1}}
          \hspace{-15pt}
          \epsfxsize=160pt
          {\epsfbox{#2}}}
}

\begin{document}

\vskip -3cm\rightline{SWAT/97/149}
\title{MONTE CARLO STUDY OF THE 3D THIRRING MODEL}

\author{ Simon HANDS (UKQCD Collaboration)}

\address{Department of Physics, University of Wales Swansea,\\
Singleton Park, Swansea SA2 8PP, UK}


\maketitle\abstracts{
I review three different non-perturbative approaches to the three
dimensional $N_f$-flavor Thirring model: the $1/N_f$ expansion, Schwinger-Dyson
equations, and Monte Carlo simulation. Simulation results are
presented to support the existence of a non-perturbative fixed point 
at a chiral symmetry breaking phase transition
for $N_f=2$ and 4, but not for $N_f=6$. Spectrum calculations 
for $N_f=2$ reveal 
conventional level ordering near the transition.}

\section{Introduction}

In this talk I will present work done in collaboration with Luigi Del
Debbio and John Mehegan \cite{us}. The Thirring model is a field theory
of relativistic fermions interacting via a contact term between
conserved currents. Its Lagrangian is written
\begin{equation}
{\cal L}=\bar\psi_i(\partial{\!\!\! /}\,+m)\psi_i
+{g^2\over{2N_f}}(\bar\psi_i\gamma_\mu\psi_i)^2,
\label{eq:Lff}
\end{equation}
where $\psi_i,\bar\psi_i$ are taken to be 4 component spinors with
flavor index $i$ running from 1 to $N_f$. Why is this an interesting
model to study in three spacetime dimensions? Firstly, it appears that
the expansion in powers of $1/N_f$ is exactly renormalisable for
$2<d<4$, which in turn suggests an equivalence between the strong
coupling limit of the Thirring model and the infra-red behaviour of 
$\mbox{QED}_3$. This result is essentially a perturbative one,
albeit using a different expansion parameter; however
non-perturbative studies using e.g. the Schwinger-Dyson approach suggest
the model may exhibit dynamical symmetry breaking and have a non-trivial
ultra-violet fixed point, of the type discussed by Miransky in the 
context of quenched $\mbox{QED}_4$ \cite{Mir}, for $N_f$ less than some
critical $N_{fc}<\infty$. At the proposed fixed point there would be 
no small dimensionless parameter in play; this is therefore a natural
candidate for investigation by Monte Carlo simulation of the
lattice-regularised model. In the following I will review each of these
approaches, before presenting our results. Finally, it is worth
mentioning that non-trivial fixed-point behaviour of $\mbox{QED}_3$ has
been proposed to account for non-Fermi liquid behaviour in high $T_c$
superconductors \cite{Mavro}.

\section{Three non-perturbative approaches}
\subsection{$1/N_f$ expansion}

The Lagrangian (\ref{eq:Lff}) may be rewritten using a vector auxiliary
field $A_\mu$:
\begin{equation}
{\cal L}=\bar\psi_i(\partial{\!\!\! /}\,+m)\psi_i
+{{ig}\over\surd{N_f}}A_\mu\bar\psi_i\gamma_\mu\psi_i+{1\over2}A_\mu
A_\mu.
\label{eq:Laux}
\end{equation}
In weak coupling perturbation theory, the vector propagator
$D_{\mu\nu}(k)$ is simply $\delta_{\mu\nu}$; however to leading order in
$1/N_f$ the vacuum polarisation bubble also contributes, with the result
that $D_{\mu\nu}$ acquires momentum dependence \cite{Paris} \cite{Gomes}:
\begin{equation}
D_{\mu\nu}(k)=\left(\delta_{\mu\nu}-{{k_\mu k_\nu}\over{k^2}}\right)
{1\over{1-\Pi(k^2)}}+\mbox{longitudinal},
\label{eq:Dmunu}
\end{equation}
with
\begin{equation}
\Pi(k^2)=-{g^2\over{2\pi}}\left(m+{1\over{2\surd{k^2}}}(k^2-4m^2)
\tan^{-1}\left({\surd{k^2}\over{2m}}\right)\right).
\label{eq:vacpol}
\end{equation}

This form can be examined in two limits. In the UV r\'egime
\begin{equation}
\displaystyle\lim_{k^2\to\infty}D_{\mu\nu}(k^2)=
\left(\delta_{\mu\nu}-{{k_\mu k_\nu}\over{k^2}}\right){8\over
{g^2\surd{k^2}}}.
\end{equation}
The $1/k$ asymptotic behaviour has the effect of softening the contact
term and thus decreasing the superficial degree of divergence associated
with higher order diagrams containing internal vector lines, such
as fermion self-energy, two-loop vacuum polarisation, etc. These
divergences can be absorbed by a finite number of counterterms,
essentially due to a novel application of Weinberg's
theorem, rendering the $1/N_f$ expansion renormalisable \cite{Hands}.
At lowest non-trivial order, the corrections coincide with those
calculated in the IR limit of $\mbox{QED}_3$~\cite{Espriu}, giving the
first hint that the two models may be related.

In the low-momentum r\'egime, (\ref{eq:Dmunu},\ref{eq:vacpol}) 
can be used to find a
pole condition and hence calculate the mass $M_V$ of the vector bound
state. The result depends on the dimensionless combination $(mg^2)$
\cite{Hands}:
\begin{equation}
M_V=\cases {2m\left(1-2\exp\left(-{{2\pi}\over{mg^2}}\right)\right)
&$mg^2\to0$;\cr
m\sqrt{{6\over{mg^2}}}&$mg^2\to\infty$.\cr}
\end{equation}
Thus we identify a smooth crossover from a 
weakly coupled theory in which the vector is a weakly
bound $f\bar f$ state to a strongly coupled one in which it is almost
massless. To leading order in $1/N_f$ the $(\bar\psi\gamma_\mu\psi)^2$
interaction is a marginal operator. In the strong coupling limit
$g^2\to\infty$ a massless vector mediates interaction between conserved
currents, suggesting that in this limit the theory is identical
to $\mbox{QED}_3$.

\subsection{Schwinger-Dyson equations}

The Schwinger-Dyson approach~\cite{Gomes}
 is to solve for the full fermion propagator
$S_F^{-1}(k)=A(k)ik{\!\!\! /}\,+\Sigma(k)$ self-consistently, by finding
integral equations for the functions $A$ and $\Sigma$. The
most systematic analysis of the Thirring model~\cite{Itoh} exploits a 
hidden local symmetry of (\ref{eq:Laux}) to fix a non-local gauge in
which the wavefunction renormalisation $A(k)\equiv1$. The usual
truncations of setting the full vector propagator to its leading order $1/N_f$
form (\ref{eq:Dmunu}) and the full vertex $\Gamma_{\mu}(p,q)$ to the
bare vertex $ig\gamma_\mu/\sqrt{N_f}$ (i.e. the ladder approximation)
are also employed. The resulting SD equation for $\Sigma(k)$ can be
solved analytically in the limit $g^2\to\infty$.

In the chiral limit $m\to0$, a non-trivial solution for $\Sigma(k)$ is
found for small values of $N_f$, signalled by a non-vanishing order
parameter, the chiral condensate $\langle\bar\psi\psi\rangle$:
\begin{equation}
\langle\bar\psi\psi\rangle\propto\Lambda^2\exp(
-3\pi/\sqrt{\textstyle{N_f\over N_{fc}}-1}),
\end{equation}
with $\Lambda$ a UV cutoff. This solution exists for
$N_f<N_{fc}=128/3\pi^2\simeq4.3$, implying that in the strong coupling
limit, a separation of IR and UV scales allowing a continuum limit 
can only be made as $N_f\to N_{fc}$. The essentially singular behaviour is
reminiscent of quenched $\mbox{QED}_4$~\cite{Mir}. For finite $g^2$
there is no analytic solution, but it has been argued that the critical
curve $N_{fc}(g)$ is smooth~\cite{Kondo}, implying that non-trivial UV
fixed points as $g^2\to g_c^2(N_f)$ may exist for integer $N_f<N_{fc}$.
The Thirring interaction has thus become relevant.
It should be noted, however, that a different truncation procedure
yields $N_{fc}=\infty$~\cite{Hong}. It is therefore desirable to try an
alternative non-perturbative approach to see if fixed points can be
identified, and to get an independent estimate of $N_{fc}$.

\subsection{Lattice formulation}

The lattice action we have used is
\bea
        S &=& \oh \sum_{x\mu i}
 \bar\chi_i(x) \eta_\mu(x) \left[\chi_i(x+\hat\mu) -
        \chi_i(x-\hat\mu)\right] + \nonumber \\
          & & m \sum_{xi} \bar\chi_i(x) \chi_i(x) +
\frac{g^2}{2N} \sum_{x\mu ij} \bar\chi_i(x) \chi_i(x+\hat\mu)
        \bar\chi_j(x+\hat\mu) \chi_j(x)
\label{eq:lattice-thirring}
\eea
where $i$ runs over $N$ flavors of one-component staggered fermion
fields $\chi,\bar\chi$, and $\eta_\mu(x)$ are the space-dependent
Kawamoto-Smit phases which are lattice analogues of the $\gamma$-matrices.
We have simulated the models with $N=1,2$ and 3, proceeding by
introducing a vector auxiliary on lattice links by analogy with
(\ref{eq:Laux}). Alternative formulations are possible~\cite{Kim}; we
have chosen (\ref{eq:lattice-thirring}) so that there is no gauge
symmetry, and the highest interaction is four-point.

There are two aspects of the formulation (\ref{eq:lattice-thirring})
which require further discussion. Firstly, and slightly unexpectedly,
the lattice interaction current is not exactly conserved, because it
couples to $iA_\mu$ and not $\exp(iA_\mu)$. This means that the lattice
vacuum polarisation tensor is not transverse, which at leading order in
$1/N_f$ results in an additive renormalisation of the inverse coupling
$(1/g^2)$~\cite{us}:
\begin{equation}
g^2\mapsto g^2_R={g^2\over{1-{2\over3}g^2}}.
\end{equation}
We should thus expect to find the strong coupling limit at
$g_R^2\to\infty$, i.e. $1/g^2\to2/3$ (for $1/g^2<2/3$ the model is not
unitary). 
It is clear that for small $N_f$,
identifying the limit $g_R^2\to\infty$ numerically will be difficult in
practice.

\begin{figure*}
\epsfaxhax{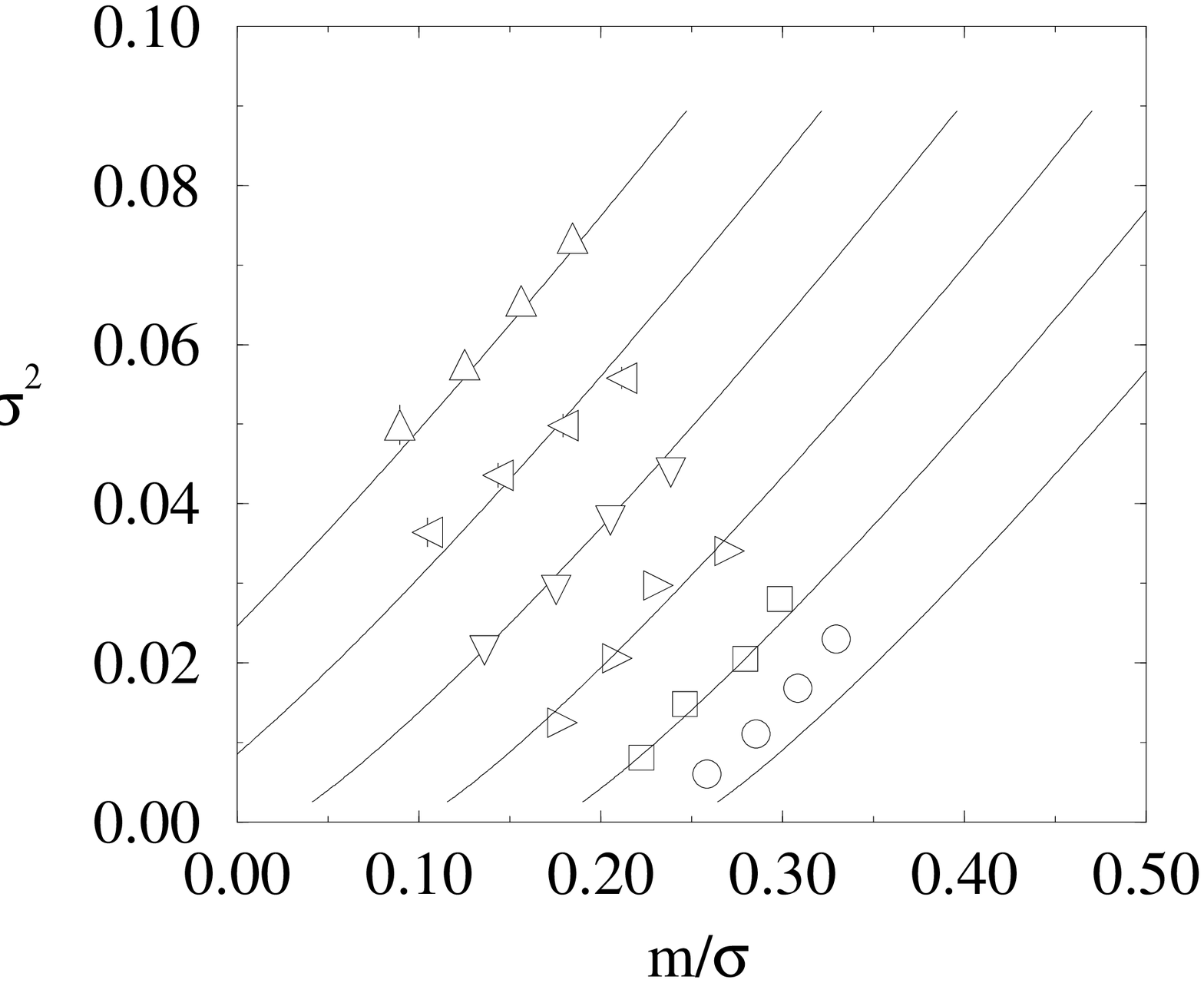}{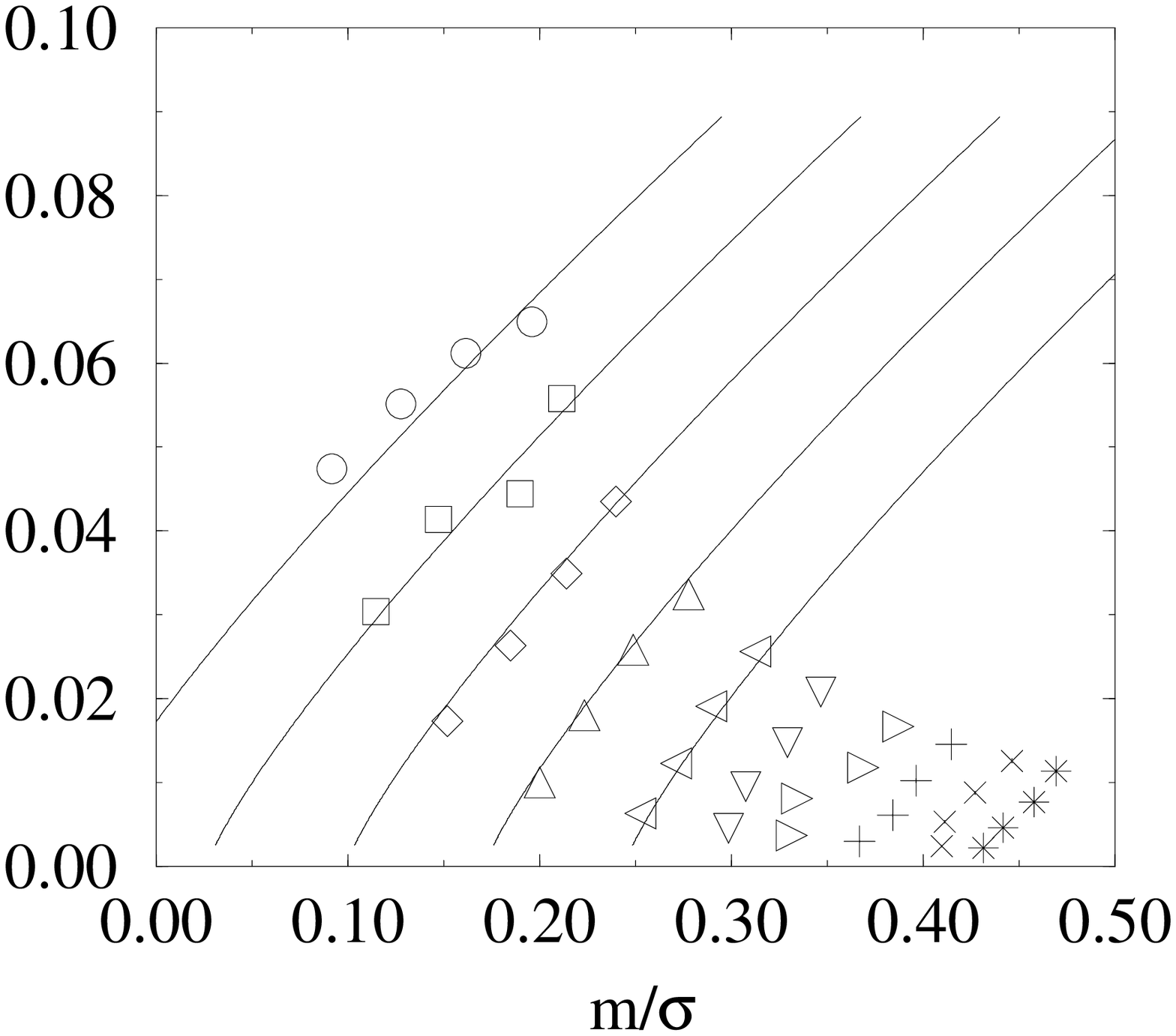}
\caption{\hskip 4.1cm Figure 2:}
\end{figure*}

Secondly, when the lattice action (\ref{eq:lattice-thirring}) is
recast in terms of physical fields $q,\bar q$ which have explicit spin
and flavor indices, then we find~\cite{us}

\noindent{\em(i)\/} Each flavor of $\chi$ corresponds to 2 physical
flavors, via fermion doubling, i.e.
\begin{equation}
N_f=2N.
\end{equation}

\noindent{\em(ii)\/} In addition to the expected interaction term
$(\bar q_i(\gamma_\mu\otimes\One)q_i)^2$, where the second $2\times2$
matrix in the product acts on flavor indices,
(\ref{eq:lattice-thirring}) contains other interactions such as 
$(\bar q_i(\gamma_5\gamma_\mu\otimes\One)q_i)^2$ and even non-covariant ones
such as $(\bar q_i(\gamma_4\gamma_\nu\gamma_\mu\otimes\tau_\nu^*)q_i)^2$.
These undesired terms are not suppressed in the 
limit lattice spacing $a\to0$. Since a strongly-coupled theory requires
a non-perturbative regularisation, we must take a robust attitude to
this problem, and determine the fixed point behaviour, if any, by
numerical measurement of critical scaling and spectral quantities.

\section{Numerical results}
\subsection{Critical scaling}

We have Monte Carlo data from lattice sizes $8^3$, $12^3$ and $16^3$,
and a range of bare masses $ma=0.05,\ldots,0.01$~\cite{us}. To
determine the critical properties requires an extrapolation to the chiral
limit $m\to0$; this is best done globally, so in the neighbourhood of
the chiral phase transition we assume an equation of state of the form
\begin{equation}
m=A\left({1\over g^2}-{1\over g_c^2}\right)\langle\bar\psi\psi\rangle^
{\delta-1/\beta}+B\langle\bar\psi\psi\rangle^\delta.
\label{eq:eos}
\end{equation}
Here $\delta$ and $\beta$ are standard critical exponents, which in a
mean field treatment would assume the values $\delta=3$,
$\beta={1\over2}$. It follows that in the mean-field case a plot of
$\langle\bar\psi\psi\rangle^2$ vs. $m/\langle\bar\psi\psi\rangle$,
known as a Fisher plot, would yield straight lines for constant $1/g^2$
which would pass through the origin at the critical coupling, and above
or below the origin in the super- or sub-critical region respectively.
In Figs. 1 and 2 we show Fisher plots for $N_f=2$ in the neighbourhood
of $1/g_c^2\simeq1.9$ and $N_f=4$ near $1/g_c^2\simeq0.6$. The lines
denote our best fits to (\ref{eq:eos}) assuming $\delta-1/\beta=1$,
which follows from the ladder approximation~\cite{DHKK}.
The curvature indicates a
slight departure from mean-field scaling. Our fits, which include a
finite volume scaling analysis for $N_f=2$, are presented in 
Tab.~\ref{tab:fit}
\begin{table}[t]
\setlength{\tabcolsep}{1.5pc}
\caption{Fits to the Equation of State}
\label{tab:fit}
\vspace{0.4cm}
\begin{center}
\begin{tabular}{|ccc|}
\hline
           &   $N_f=2$   &   $N_f=4$    \\
\hline
$1/g_c^2$  &  1.92(2)    &  0.66(1)     \\
$\delta$   &  2.75(9)    &  3.43(9)     \\
\hline
\end{tabular}
\end{center}
\end{table}

The remaining exponents can be derived using the constraint between
$\delta$ and $\beta$, and hyperscaling. The most significant result
is not so much that the value of $\delta$ departs from the mean-field
value in each case, but rather that the values from each model are
distinct. This is consistent with the $N_f=2,4$ theories belonging to
different universality classes.

A Fisher plot for $N_f=6$, shown in Fig. 3, shows no evidence for a
chirally broken phase, and no fit of the form (\ref{eq:eos}) is
possible. Thus the data, from admittedly small lattices to date,
strongly suggest
\begin{equation}
N_{fc}<6.
\end{equation}

\subsection{Spectroscopy}

We have measured the physical fermion and bound state masses for
$N_f=2$ on a $16^3$ lattice with $ma=0.01$. Fig. 4 shows fermion, pion
and scalar masses vs. $1/g^2$, showing a sharp crossover near
$1/g^2_c\simeq1.9$. In the symmetric phase both pion and scalar look
like weakly bound $f\bar f$ states, while in the broken phase the pion
is the lightest particle, consistent with Goldstone's theorem. There is
no evidence for the absence of light bound states in the symmetric
phase, predicted if the fixed point is described by an essential
singularity~\cite{MirYam}. We have also examined the vector channel,
where the simplest interpolating operator projects onto states with 
both vector ($\bar q\gamma_\mu q$) and axial vector ($\bar
q\gamma_5\gamma_\mu q$) quantum numbers. Surprisingly, we found both
states to be light in the symmetric phase, even though to leading order
in $1/N_f$ there is no bound state in the axial vector channel. This is
a hint that the $1/N_f$ expansion may be failing to describe the physics
even of the symmetric phase near the fixed point. Measurements in the
broken phase are much noisier, but there is tentative evidence that the
vector mass increases sharply.

\begin{figure*}
\epsfaxhax{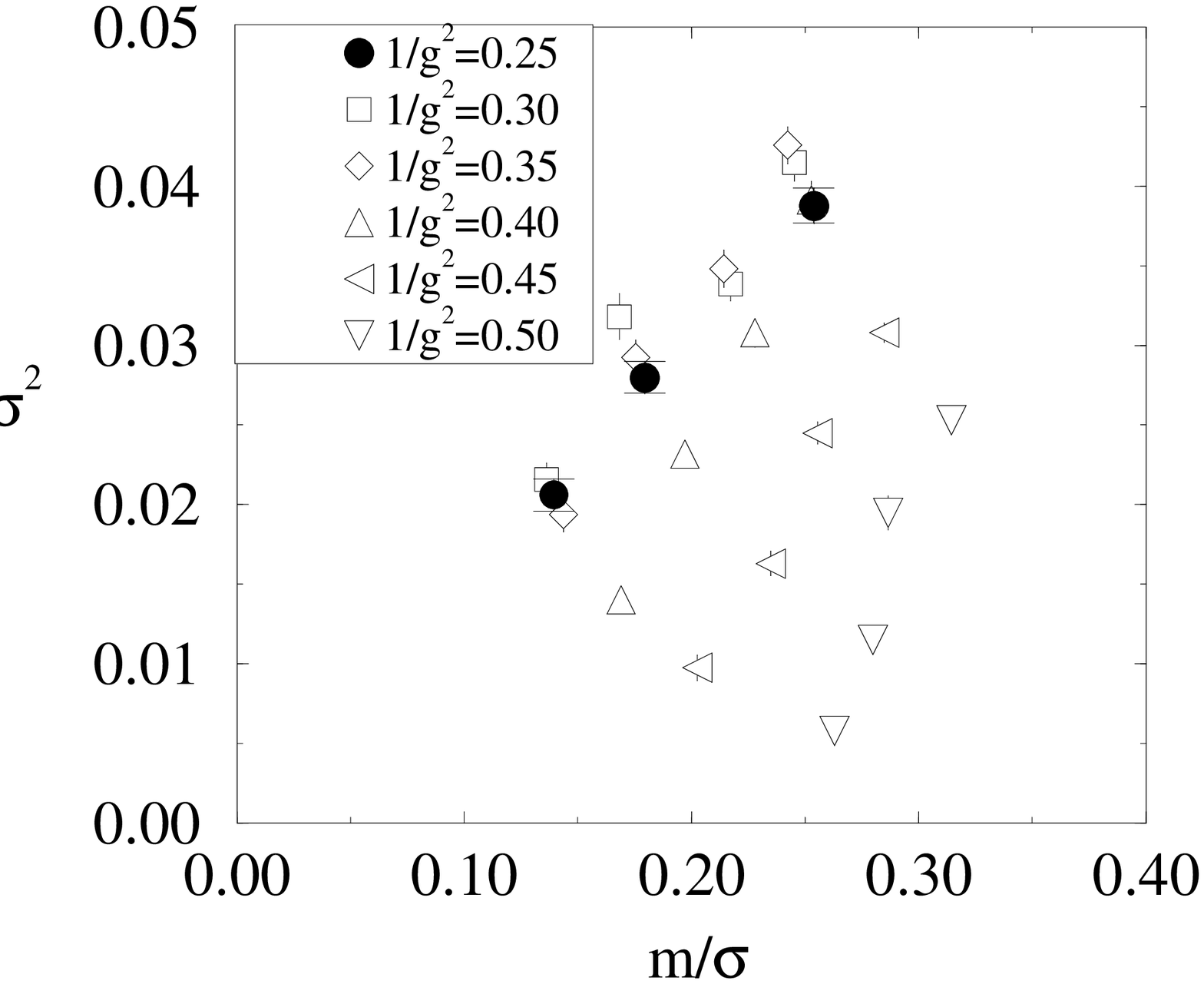}{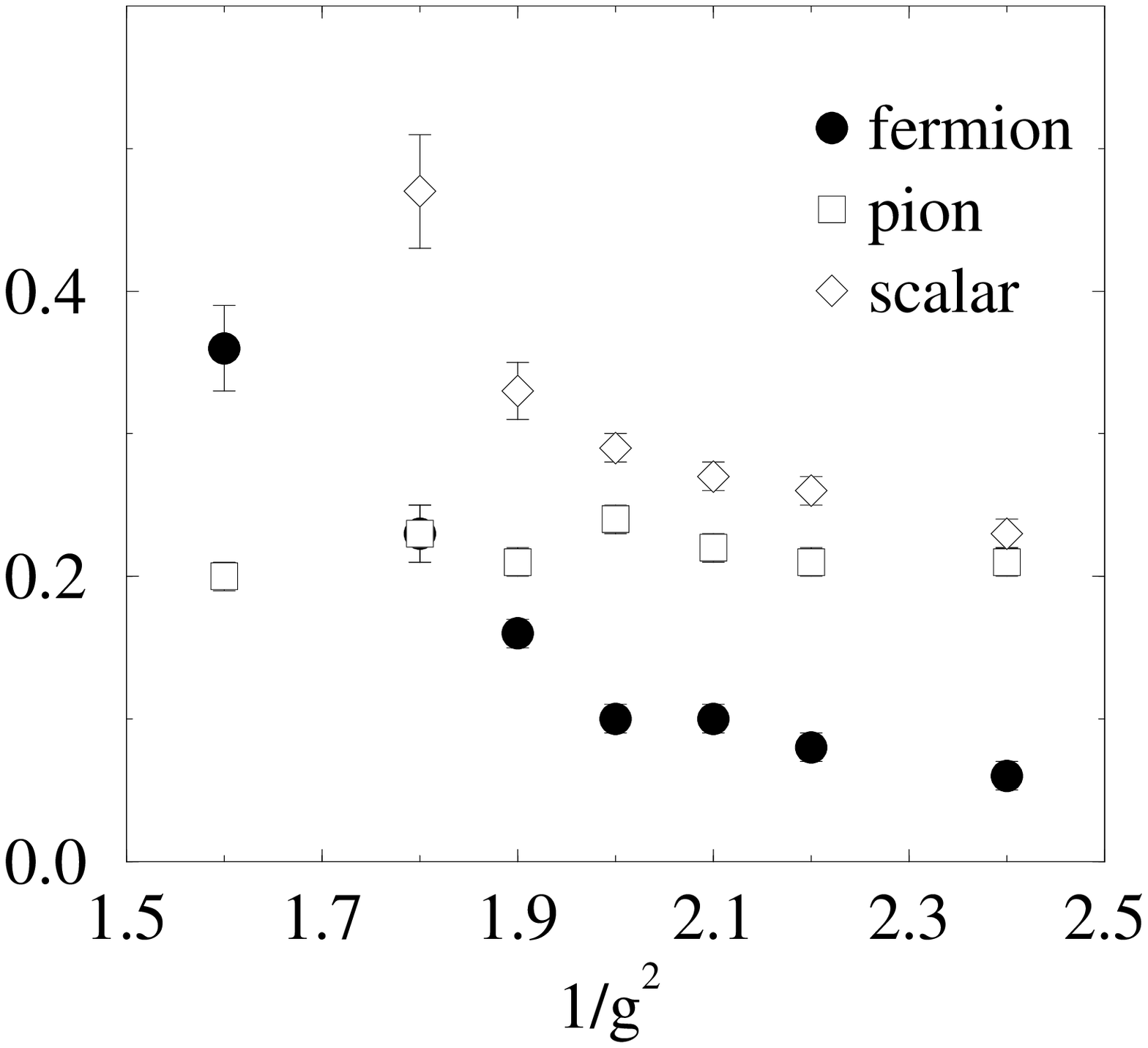}
\setcounter{figure}{2}
\caption{\hskip 4.1cm Figure 4:}
\end{figure*}

\section{Summary}

Our main results are that for $N_f=2,4$ there is clear evidence for a
continuous chiral phase transition, with scaling described by distinct
critical exponents in each case. No such transition is seen for $N_f=6$,
implying $N_{fc}<6$. For $N_f=2$, we see conventional level ordering of
the spectrum in fermion, pion and scalar channels across the transition,
and light states in the symmetric phase in both vector and axial vector
channels.

\section*{Acknowledgments}
The author is supported by a PPARC Advanced Research Fellowship, with
additional travel funding from the Royal Society. Some of
the computational work was performed using resources granted under 
PPARC grants GR/J67475, GR/K41663, GR/K455745 and GR/L29927.

\end{document}